\documentclass[useAMS,usenatbib]{mn2e}
\usepackage{graphicx}
\title{Cosmic slowing down of acceleration using $f_{gas}$ }

\author[V\'ictor H. Cardenas, Carla Bernal and Alexander Bonilla]{V\'ictor H. C\'ardenas\thanks{E-mail: victor.cardenas[at]uv.cl},
Carla Bernal\thanks{E-mail: carla.bernalb[at]alumnos.uv.cl} and
Alexander Bonilla\thanks{E-mail: alexander.bonilla[at]uv.cl}\\
Departamento de F\'{\i}sica y Astronom\'ia, Facultal de Ciencias,
Universidad de Valpara\'iso,\\ Gran Breta\~na 1111, Valpara\'iso,
Chile \\ Centro de Astrof\'isica de Valpara\'iso, CAV, Gran
Breta\~na 1111, Valpara\'iso, Chile}

\begin{document}


\pagerange{\pageref{firstpage}--\pageref{lastpage}} \pubyear{2002}

\maketitle

\label{firstpage}

\begin{abstract}
We investigate the recent - low redshift - expansion history of the
universe using the most recent observational data. Using only data
from 42 measurements of $f_{gas}$ in clusters, we found that cosmic
acceleration could have already peaked and we are witnessing now its
slowing down. This effect, found previously by Shafieloo, Sahni and
Starobinsky in 2010 using supernova data (at that time the
Constitution SNIa sample) appears again using an independent
observational probe. We also discuss the result using the most
recent Union 2.1 data set.
\end{abstract}


\section{Introduction}

Dark energy is the name of the mysterious component which drives the
current accelerated expansion of the universe \citep{dereview}. In
its simplest form, this can be described by a fluid with constant
equation of state parameter $w=-1$, which corresponds to a
cosmological constant, leading to the successful $\Lambda$CDM model;
the simplest model that fits a varied set of observational data.

This model has several unnatural properties, that lead us look for
models whose description is less forced. For example, the current
value for $\Omega_{\Lambda}$ and $\Omega_M$ are of the same order of
magnitude, a fact highly improbable, because the dark matter
contribution decreases with $a^{-3}$, where $a(t)$ is the scale
factor, meanwhile the cosmological constant contribution have had
always the same value. This problem in particular is known as the
cosmic coincidence problem.

Thus, it seems necessary to assume the existence of a dynamical
cosmological constant, or a theoretical model with a dynamical
equation of state parameter $w(z)=p/\rho$. The source of this
dynamical \textit{dark energy} could be either, a new field
component filling the universe, as a quintessence scalar field
\citep{quinta1,quinta2,quinta3,quinta4,quinta5}, or it can be
produced by modifying gravity \citep{Tsujikawa2010},
\citep{Capozziello2011}, \citep{Starkman2011}.

In \citep{shafi2009} the authors suggested the current observational
data favor a scenario in which the acceleration of the expansion has
past a maximum value and is now decelerating. The key point in
deriving this conclusion is the use of the Chevalier-Polarski-Linder
(CPL) parametrization \citep{Chevallier:2000qy},
\citep{Linder:2002et} for a dynamical equation of state parameter
\begin{equation}\label{cpl}
 w(a)=w_0+(1-a)w_1,
\end{equation}
where $w_0$ and $w_1$ are constants to be fixed by observations, in
that case the Constitution data set \citep{constitution}. Using the
Union 2 data set \citep{Union2} the authors in \citep{Li:2010da}
found similar conclusions, under the assumption of a flat universe.
In \citep{Cardenas:2011a}, we revisit this problem using the Union 2
data set and we extend the analysis to curved spacetimes. We found
the three observational test, SNIa, BAO and CMB, can all be
accommodated in the same trend, assuming a very small value for the
curvature parameter, $\Omega_k \simeq -0.08$. The best fit values
found suggested the acceleration of the universe has already reached
its maximum, and is currently moving towards a decelerating phase.

In this paper we investigate the recent (low redshift) expansion
history of the universe in light of recent data. We use the gas mass
fraction in clusters - 42 measurements of $f_{gas}$ extracted from
\citep{Allen:2007ue} - and the latest Union 2.1 data set
\citep{Suzuki:2011hu} of supernovae type Ia (SNIa). We also consider
the constraints from baryon acoustic oscillation (BAO)
\citep{bao2,bao3} through the distance scale, and also from cosmic
background radiation (CMB) from the WMAP 9 years distance posterior
\citep{cmb2}.

The paper is organized as follows: in the next section we re-study
the flat universe case using both SNIa data sets
\citep{constitution}, and \citep{Union2}. Then we present the
results using the $f_{gas}$ data. After that, we discuss the result
using the latest Union 2.1 SNIa data. We end with a discussion of
the results.

\section{$q(z)$ at low redshift from SNIa}

In this section we describe the results obtained by
\citep{shafi2009} and \citep{Li:2010da} for the deceleration
parameter at small redshift using the Constitution
\citep{constitution} and Union 2 \citep{Union2} data sets. The
analysis of this section is restricted to flat universes. The
extension to curved spacetime was performed first in
\citep{Cardenas:2011a}, and can be implemented directly.

The first step is to define a cosmological model to test. The
comoving distance from the observer to redshift $z$ is given by
\begin{equation}\label{comdistance}
r(z) =  \frac{c}{H_0} \int_0^z \frac{dz'}{E(z')},
\end{equation}
where
\begin{equation}\label{edez}
E^2(z)  =  \Omega_m (1+z)^3+\Omega_{de}f(z),
\end{equation}
where $\Omega_{de}=1-\Omega_m$ and in general
\begin{equation}\label{fdez}
  f(z)  =  \exp \left\{ 3 \int^z_0 \frac{1+w(z')}{1+z'} dz'
  \right\}.
\end{equation}
In this case, the cosmological model is defined through a selection
of the equation of state parameter $w(z)$. Using the CPL
parametrization (\ref{cpl}), $w=w_0+w_1 z/(1+z)$, the function
defined in (\ref{fdez}) takes the form
\begin{equation}
f(z)=(1+z)^{3(1+w_0+w_1)}\exp \left(-\frac{3w_1z}{1+z}\right).
\end{equation}
Performing the analysis, we find the best values of the parameters
that fit the observations (the details of the statistical analysis
is described in Appendix A), which in turns give us the best Hubble
function $E(z)$ as a function of redshift that agrees with the data.
From it, we can compute the deceleration parameter as
\begin{equation}\label{qdzeq}
q(z)=-(1+z)\frac{1}{E(z)}\frac{dE(z)}{dz}-1.
\end{equation}
Using the Constitution sample consisting of $397$ SNIa data points,
leads to the results shown in Table \ref{tab:table0}, which are very
similar to the ones informed in \citep{shafi2009} and
\citep{Wei:2009ry}.
\begin{table}
\centering \caption{\label{tab:table0} The best fit values for the
free parameters using the Constitution data set in the case of a
flat universe model using only SNIa data (1), SNIa+BAO (2) and
SNIa+BAO+CMB (3). See the related Fig. \ref{fig0}.}
\begin{tabular}{ccccc}
Set & $\chi^2_{min}/dof$ & $\Omega_m$ & $w_0$ & $w_1$ \\
\hline
(1) & 461.2/394 & 0.452$\pm$ 0.043 & -0.22$\pm$ 0.59 & -11$\pm$7 \\
(2) & 466.6/400 & 0.303$\pm$ 0.034 & -0.89$\pm$ 0.25 & -1$\pm$2 \\
(3) & 468.2/403 & 0.281$\pm$ 0.014 & -0.99$\pm$ 0.12 & -0.1$\pm$0.5  \\
\end{tabular}
\end{table}
\begin{figure}
  \begin{center}
    \includegraphics[width=7cm]{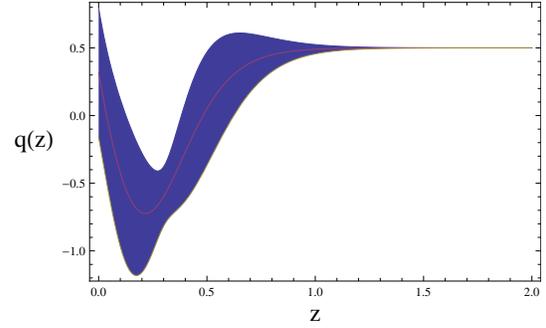}
  \end{center}
 \caption{\small Using the Constitution data set
\citep{constitution} we plot the deceleration parameter
reconstructed using the best fit values for the SNIa case. We
consider the error propagation at one sigma in the best fit
parameters.}
  \label{fig0}
\end{figure}
In particular, using only the SNIa data, we plot the deceleration
parameter as a function of redshift in Fig.(\ref{fig0}). We observe
that even at one sigma, the expansion of the universe peaks at
redshift $z\simeq 0.2$ and then evolves towards a non decelerating
regime today. If we add the BAO constraints, this behavior is still
observed. However, as was mentioned in the original papers,
incorporating the CMB constraints leads to a smooth deceleration
acceleration transition for $q(z)$, typically of $\Lambda$CDM (a
$w_0 \simeq -1$ and $w_1 \simeq 0$). The values of $\chi^2_{min}$
show that something is wrong with the CPL parametrization in trying
to fit the three observational probes. The large negative value of
$w_1$ is the responsible of this anomalously low redshift behavior
using supernovae shown in Figure \ref{fig0}. Although the error in
the determination of $w_1$ is larger using only SNIa compared to the
other cases, even at one sigma the low redshift transition is still
observable.

Using the Union 2 set \citep{Union2} consisting in $557$ SNIa, the
analysis leads to the results shown in Table \ref{tab:table01}.
\begin{table}
\caption{\label{tab:table01} The best fit values for the free
parameters using the Union 2 data set in the case of a flat universe
model using only SNIa data (1), SNIa+BAO (2) and SNIa+BAO+CMB (3).
See also Fig. \ref{fig01}.}
\begin{tabular}{ccccc}
Set & $\chi^2_{min}/dof$ & $\Omega_m$ & $w_0$ & $w_1$ \\
\hline
(1) & 541.43/554 & 0.420$\pm$0.068 & -0.86$\pm$0.38 & -5.5$\pm$5.4 \\
(2) & 542.11/560 & 0.294$\pm$0.033 & -1.01$\pm$0.15 & -0.4$\pm$1.2  \\
(3) & 543.91/563 & 0.280$\pm$0.015 & -1.07$\pm$0.12 & 0.23$\pm$0.37 \\
\end{tabular}
\end{table}
\begin{figure}
\begin{center}
\includegraphics[width=7cm]{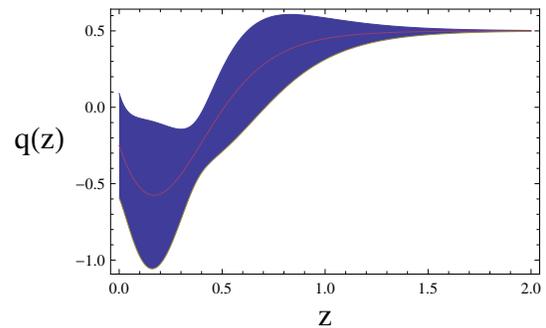}
\end{center}
\caption{ Using the Union 2 data set \citep{Union2} we plot the
deceleration parameter reconstructed using the best fit values for
the SNIa case. We consider the error propagation at one sigma in the
best fit parameters.}\label{fig01}
 \end{figure}
Again, we found large negative values for $w_1$ in the SNIa and
SNIa+BAO fit, but in the case with CMB data, the value of this
parameter appears smaller and positive.

This feature -- the inversion of $q(z)$ at $z\simeq 0.2$ -- can be
interpreted in two ways: first, we can assume the origin of the
feature is the use of an unappropriated parametrization, in this
case we can say that the CPL parametrization is inadequate to fit
data at large and small redshift. Clearly, a large negative value
for $w_1$ spoils the large $z$ behavior of $w(z)$. The other way, is
to consider the data from SNIa and also from BAO are telling us
something about our recent - low redshift - expansion history of the
universe, which the CMB data cannot detect.

This is the kind of small redshift transitions in $w(z)$ that were
discussed first in \citep{Bassett:2002qu} and also by
\citep{Mortonson:2009qq}. In \citep{Bassett:2002qu} they used a
form,
\begin{equation}\label{w2}
w(z)=w_0 + \frac{w_f - w_0}{1+\exp((z - z_t)/\Delta)},
\end{equation}
that captures the essence of a single transition at $z_t$ in a range
$\Delta$, from $w_0$ initially to a final value $w_f$ in the future.
In \citep{Mortonson:2009qq} the authors discussed the possibility of
a fast change in $w(z)$ at $z<0.02$ and its implication for a
standard scalar field model. A more recent work focused on small
redshift transitions in parameterizations of $q(z)$ instead of
$w(z)$ \citep{qdzparam1,qdzparam2,qdzparam3}, the so called
cosmographic approach. Even some authors have claimed the universe
has already entered into a decelerated phase using only low redshift
($z<0.1$) SNIa data \citep{lowzsnia}.

To settle this dilemma, independent evidence is required to identify
whether this effect is real or not. In the next section we study
this problem using data from gas mass fraction in clusters, an
observational probe completely independent of the SNIa.

\section{Results using the $f_{gas}$ data}

Sasaki \citep{Sasaki:1996zz} was the first who described the method
of using measurements of the apparent dependence on redshift of the
baryonic mass fraction, to constrain cosmological parameters. The
method is based on theoretical arguments and simulation results
indicating that this mass fraction, in the largest clusters, must be
constant with redshift $z$ \citep{eke98}. This null evolution with
$z$ it is only possible if the reference cosmology used in making
the measurements of the baryonic mass fraction match the true
cosmology.

In this section we use the data from \citep{Allen:2007ue} which
consist in 42 measurements of the X-ray gas mass fraction $f_{gas}$
in relaxed galaxy clusters spanning the redshift range $0.05<z<1.1$.
The $f_{gas}$ data are quoted for a flat $\Lambda$CDM reference
cosmology with $h=H_0/100$ km s$^{-1}$Mpc$^{-1}=0.7$ and
$\Omega_M=0.3$.

To determine constraints on cosmological parameters we use the model
function \citep{Allen:2004cd}
\begin{equation}\label{fgas}
f_{gas}^{\Lambda CDM}(z)=\frac{b \Omega_b}{(1+0.19\sqrt{h})
\Omega_M} \left[\frac{d_A^{\Lambda CDM}(z)}{d_A(z)} \right]^{3/2},
\end{equation}
where $b$ is a bias factor which accounts that the baryon fraction
is slightly lower than for the universe as a whole. From
\citep{eke98} it is obtained $b=0.824 \pm 0.0033$. In the analysis
we also use standard Gaussian priors on $\Omega_b h^2 = 0.0214 \pm
0.0020$ \citep{Kirkman:2003uv} and $h=0.72 \pm 0.08$
\citep{Freedman:2000cf}.

The results of the analysis are shown in Table \ref{tab:table02},
where we have omitted the best fit values for $\Omega_b$, $b$ and
$h$ because they do not show significant variations.
\begin{table}
\caption{\label{tab:table02} The best fit values for the free
parameters using only $f_{gas}$ data (1), $f_{gas}$+BAO (2) and
$f_{gas}$+BAO+CMB (3) in the case of a flat universe model. See also
Fig. \ref{fig02}. We have omitted the best fit values for
$\Omega_b$, $b$ and $h$ because they do not show significant
variations.}
\begin{tabular}{ccccc}
Data Set & $\chi^2_{min}/dof$ & $\Omega_m$ & $w_0$ & $w_1$ \\
\hline
(1) & 41.63/39 & 0.271$\pm$0.019 & -0.83$\pm$0.65 & -2.4$\pm$5.8 \\
(2) & 43.40/45 & 0.280$\pm0.016$ & -0.68$\pm$0.41 & -3.3$\pm$3.2  \\
(3) & 43.91/48 & 0.270$\pm$0.010 & -1.34$\pm$0.12 & 1.0$\pm$0.4 \\
\end{tabular}
\end{table}
\begin{figure}
\begin{center}
\includegraphics[width=7cm]{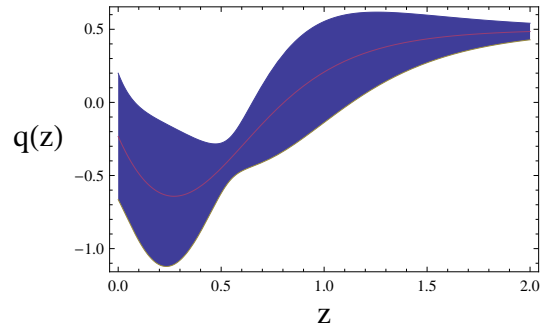}
\end{center}
\caption{ Using the $f_{gas}$ data from \citep{Allen:2007ue} we plot
the deceleration parameter reconstructed using the best fit values
for the $f_{gas}$+BAO case. We consider the error propagation at one
sigma in the best fit parameters.}\label{fig02}
 \end{figure}
In Fig.(\ref{fig02}) we plot the decelerated parameter as a function
of redshift in the case of $f_{gas}$+BAO. We get the same trend at
small $z$ previously found using the Constitution and Union 2
samples of SNIa within one sigma. This is an independent
confirmation that the data is telling us something about our local
universe.

\section{$q(z)$ from the Union 2.1 data alone}

\begin{figure}
\begin{center}
\includegraphics[width=7cm]{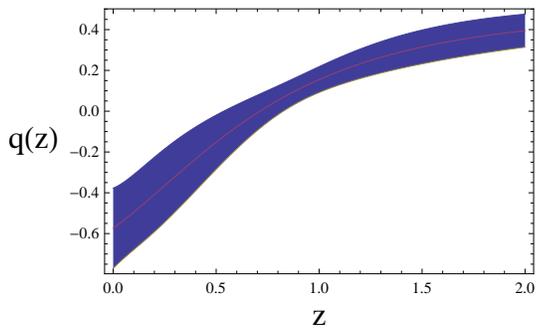}
\end{center}
\caption{ Using the Union 2.1 data from \citep{Suzuki:2011hu} we
plot the deceleration parameter reconstructed using the best fit
values for the SNIa+BAO case.} \label{fig03}
\end{figure}

The latest sample of SNIa is the Union 2.1 \citep{Suzuki:2011hu}. It
consist in 580 data points spanning a redshift range of
$0.015<z<1.41$. The results, performing the same analysis as before,
are displayed in Table \ref{tab:table03} and also in
Fig.(\ref{fig03}).
\begin{table}
\caption{\label{tab:table03} The best fit values for the free
parameters using the Union 2.1 data set in the case of a flat
universe model. See also Fig. \ref{fig03}.}
\begin{tabular}{ccccc}
Set & $\chi^2_{min}/dof$ & $\Omega_m$ & $w_0$ & $w_1$ \\
\hline
(1) & 562.22/577 & 0.27$\pm$0.47 & -1.00$\pm$0.43 & 0.$\pm$5 \\
(2) & 563.64/583 & 0.295$\pm$0.032 & -1.02$\pm$0.18 & -0.2$\pm$1.1  \\
(3) & 564.65/586 & 0.279$\pm$0.014 & -1.04$\pm$0.11 & 0.1$\pm$0.5 \\
\end{tabular}
\end{table}

Clearly, the latest SNIa data does not shown any of the features at
small redshift as the previous two data set did. As it is mentioned
explicitly in \citep{Suzuki:2011hu}, the analysis procedure used for
the Union 2 compilation is again used in the Union 2.1, with the
following change: a correction for the host-mass SNIa luminosity
relation.

The fact that this low redshift effect has been detected in the
previous SNIa data sets, and that from now on, after some correction
(from systematics) it disappears completely, can at least be
considered disturbing. This is even more intriguing considering the
findings in this paper, which points toward independent evidence of
such a low redshift behavior.

A thorough study of this low redshift behavior of SNIa will be
discussed elsewhere.

\section{Discussion}

In this paper, using $f_{gas}$ data from clusters, we have found
evidence for a low redshift transition of the deceleration
parameter, indicating that the acceleration has passed a maximum
around $z \simeq 0.2 $ and now evolves towards a decelerating phase
in the near future.

This finding is in agreement with previous studies using the
Constitution \citep{constitution} and Union 2 \citep{Union2} data
sets for SNIa, but they are in intriguing disagreement with the
latest Union 2.1 set \citep{Suzuki:2011hu}.

Our analysis then strongly suggest that low redshift measurements
have reached the precision to scrutinize the local universe, this
time not only with SNIa but also with galaxy clusters.

Regarding the intriguing disagreement mentioned, the only
significant change in the analysis performed between the Union 2.1
data set and the previous ones, was the correction for the host-mass
SNIa luminosity relation.

This is exactly the case of a probable related issue considered in
the past; the so called Hubble bubble. In \citep{Zehavi:1998gz}
using around forty SNIa, they found evidence for a local void or
Hubble bubble as was named after. In \citep{Jha:2006fm}, almost ten
years after, the authors found new evidence pointing to similar
conclusions.

Their existence has been related to the details of the treatment of
extinction and reddening by dust \citep{Conley:2007ng}. Hicken et
al., demonstrated \citep{Hicken:2009df} that no such a Hubble bubble
exist if a reduced value $R_{V}=1.7$ is used instead.

The previous work cite \citep{Jha:2006fm} used a MLCS2k2 fitting
method with a reddening parameter $R_{V}=3.1$. The Constitution and
Union 2 samples were derived using a Milky Way extinction
$R_{V}=3.1$ from \citep{Cardelli:1989fp}, meanwhile the Union 2.1
uses a lower value $R_{V}=1.7$.

As it was stressed also in \citep{Wiltshire:2012uh} there is a
danger in considering the $R_{V}$ parameter as an adjustable
parameter in the light curve reduction. Actually, independent
studies have found averages values of $R_{V}=2.77 \pm 0.41$
\citep{Finkelman:2008nm}, \citep{Finkelman:2010jt} using galaxies,
see also \citep{Motta:2002se}, \citep{mediavilla} for a lens galaxy
at $z=0.88$, a very much the Milky Way value, so it is an important
issue to take care of the treatment of the $R_{V}$ parameter in SNIa
studies.

Our results suggest to perform a careful study at small redshift to
elucidate the low redshift - apparently anomalously - behavior of
$q(z)$.

\section*{Acknowledgments}

The author want to thank Ver\'onica Motta, Juan Maga\~na and Sergio
del Campo for useful discussions. This work was funded by Comisi\'on
Nacional de Ciencias y Tecnolog\'{\i}a through FONDECYT Grant
1110230 and DIUV 13/2009.

\appendix

\section{Statistical Analysis}

The SNIa data give the luminosity distance $d_L(z)=(1+z)r(z)$. We
fit the SNIa with the cosmological model by minimizing the $\chi^2$
value defined by
\begin{equation}
\chi_{SNIa}^2=\sum_{i=1}^{557}\frac{[\mu(z_i)-\mu_{obs}(z_i)]^2}{\sigma_{\mu
i}^2},
\end{equation}
where  $\mu(z)\equiv 5\log_{10}[d_L(z)/\texttt{Mpc}]+25$ is the
theoretical value of the distance modulus, and $\mu_{obs}$ is the
corresponding observed one.

The BAO measurements considered in our analysis are obtained from
the WiggleZ experiment \citep{2011MNRAS.tmp.1598B}, the SDSS DR7 BAO
distance measurements \citep{2010MNRAS.401.2148P}, and 6dFGS BAO
data \citep{2011MNRAS.416.3017B}.

The $\chi^2$ for the WiggleZ BAO data is given by
\begin{equation}
\chi^2_{\scriptscriptstyle WiggleZ} = (\bar{A}_{obs}-\bar{A}_{th})
C_{\scriptscriptstyle WiggleZ}^{-1} (\bar{A}_{obs}-\bar{A}_{th})^T,
\end{equation}
where the data vector is $\bar{A}_{obs} = (0.474,0.442,0.424)$ for
the effective redshift $z=0.44,0.6$ and 0.73. The corresponding
theoretical value $\bar{A}_{th}$ denotes the acoustic parameter
$A(z)$ introduced by \citet{2005ApJ...633..560E}:
\begin{equation}
A(z) = \frac{D_V(z)\sqrt{\Omega_{m}H_0^2}}{cz},
\end{equation}
and the distance scale $D_V$ is defined as
\begin{equation}
D_V(z)=\frac{1}{H_0}\left[(1+z)^2D_A(z)^2\frac{cz}{E(z)}\right]^{1/3},
\end{equation}
where $D_A(z)$ is the Hubble-free angular diameter distance which
relates to the Hubble-free luminosity distance through
$D_A(z)=D_L(z)/(1+z)^2$. The inverse covariance
$C_{\scriptscriptstyle WiggleZ}^{-1}$ is given by
\begin{equation}
C_{\scriptscriptstyle WiggleZ}^{-1} = \left(
\begin{array}{ccc}
1040.3 & -807.5 & 336.8\\
-807.5 & 3720.3 & -1551.9\\
336.8 & -1551.9 & 2914.9
\end{array}\right).
\end{equation}

Similarly, for the SDSS DR7 BAO distance measurements, the $\chi^2$
can be expressed as \citep{2010MNRAS.401.2148P}
\begin{equation}
\chi^2_{\scriptscriptstyle SDSS} =
(\bar{d}_{obs}-\bar{d}_{th})C_{\scriptscriptstyle
SDSS}^{-1}(\bar{d}_{obs}-\bar{d}_{th})^T,
\end{equation}
where $\bar{d}_{obs} = (0.1905,0.1097)$ is the datapoints at $z=0.2$
and $0.35$. $\bar{d}_{th}$ denotes the distance ratio
\begin{equation}
d_z = \frac{r_s(z_d)}{D_V(z)}.
\end{equation}
Here, $r_s(z)$ is the comoving sound horizon,
\begin{equation}
 r_s(z) = c \int_z^\infty \frac{c_s(z')}{H(z')}dz',
 \end{equation}
where the sound speed $c_s(z) = 1/\sqrt{3(1+\bar{R_b}/(1+z)}$, with
$\bar{R_b} = 31500 \Omega_{b}h^2(T_{CMB}/2.7\rm{K})^{-4}$ and
$T_{CMB}$ = 2.726K.

The redshift $z_d$ at the baryon drag epoch is fitted with the
formula proposed by \citet{1998ApJ...496..605E},
\begin{equation}
z_d =
\frac{1291(\Omega_{m}h^2)^{0.251}}{1+0.659(\Omega_{m}h^2)^{0.828}}[1+b_1(\Omega_b
h^2)^{b_2}],
\end{equation}
where
\begin{eqnarray}
&b_1 = 0.313(\Omega_{m}h^2)^{-0.419}[1+0.607(\Omega_{m}h^2)^{0.674}], \\
&b_2 = 0.238(\Omega_{m}h^2)^{0.223}.
\end{eqnarray}

$C_{\scriptscriptstyle SDSS}^{-1}$ in Eq. (12) is the inverse
covariance matrix for the SDSS data set given by
\begin{equation}
C_{\scriptscriptstyle SDSS}^{-1} = \left(
\begin{array}{cc}
30124 & -17227\\
-17227 & 86977
\end{array}\right).
\end{equation}

For the 6dFGS BAO data \citep{2011MNRAS.416.3017B}, there is only
one data point at $z=0.106$, the $\chi^2$ is easy to compute:
\begin{equation}
\chi^2_{\scriptscriptstyle 6dFGS} =
\left(\frac{d_z-0.336}{0.015}\right)^2.
\end{equation}

The total $\chi^2$ for all the BAO data sets thus can be written as
\begin{equation}
\chi^2_{BAO} = \chi^2_{\scriptscriptstyle WiggleZ} +
\chi^2_{\scriptscriptstyle SDSS} + \chi^2_{\scriptscriptstyle
6dFGS}.
\end{equation}

We also include CMB information by using the WMAP 9-yr data
\citep{cmb2} to probe the expansion history up to the last
scattering surface. The $\chi^2$ for the CMB data is constructed as
\begin{equation}\label{cmbchi}
 \chi^2_{CMB} = X^TC_{CMB}^{-1}X,
\end{equation}
where
\begin{equation}
 X =\left(
 \begin{array}{c}
 l_A - 302.40 \\
 R - 1.7246 \\
 z_* - 1090.88
\end{array}\right).
\end{equation}
Here $l_A$ is the ``acoustic scale'' defined as
\begin{equation}
l_A = \frac{\pi d_L(z_*)}{(1+z)r_s(z_*)},
\end{equation}
where $d_L(z)=D_L(z)/H_0$ and the redshift of decoupling $z_*$ is
given by \citep{husugi},
\begin{equation}
z_* = 1048[1+0.00124(\Omega_b h^2)^{-0.738}]
[1+g_1(\Omega_{m}h^2)^{g_2}],
\end{equation}
\begin{equation}
g_1 = \frac{0.0783(\Omega_b h^2)^{-0.238}}{1+39.5(\Omega_b
h^2)^{0.763}},
 g_2 = \frac{0.560}{1+21.1(\Omega_b h^2)^{1.81}},
\end{equation}
The ``shift parameter'' $R$ defined as \citep{BET97}
\begin{equation}
R = \frac{\sqrt{\Omega_{m}}}{c(1+z_*)} D_L(z).
\end{equation}
$C_{CMB}^{-1}$ in Eq. (\ref{cmbchi}) is the inverse covariance
matrix,
\begin{equation}
C_{CMB}^{-1} = \left(
\begin{array}{ccc}
3.182 & 18.253 & -1.429\\
18.253 & 11887.879 & -193.808\\
-1.429 & -193.808 & 4.556
\end{array}\right).
\end{equation}

\bsp

\label{lastpage}

\end{document}